\documentclass[conference]{IEEEtran}
\IEEEoverridecommandlockouts
\usepackage{cite}
\usepackage{amsmath,amssymb,amsfonts}
\usepackage{graphicx}
\usepackage{bbding}
\usepackage{textcomp}
\usepackage{xcolor}
\usepackage{algorithm}
\usepackage{amsmath}
\usepackage[noend]{algpseudocode}
\usepackage{hyperref}
\usepackage{booktabs}
\usepackage{stfloats} 
\usepackage{multirow}
\usepackage{multicol}
\usepackage{listings}
\usepackage{subcaption}
\usepackage{setspace}
\usepackage[dvipsnames,svgnames]{xcolor}
\usepackage[many]{tcolorbox} 

\definecolor{main}{HTML}{222222}    
\definecolor{sub}{HTML}{dddddd}     

\tcbset{
    sharp corners,
    colback = white,
    before skip = 0.2cm,    
    after skip = 0.5cm      
}                           

\newtcolorbox{boxD}{
    colback = sub, 
    colframe = main, 
    boxrule = 0pt, 
    toprule = 3pt, 
    bottomrule = 3pt 
}

\hypersetup{
    colorlinks=true,
    linkcolor=black,
    citecolor=black,
    urlcolor=blue,
    pdfborder={0 0 0}
}

\def\BibTeX{{\rm B\kern-.05em{\sc i\kern-.025em b}\kern-.08em
    T\kern-.1667em\lower.7ex\hbox{E}\kern-.125emX}}
\begin{document}

\title{FOCAL: A Novel Benchmarking Technique for Multi-modal Agents 
}

 \author{\IEEEauthorblockN{Aditya Choudhary \IEEEauthorrefmark{2} }
 \IEEEauthorblockA{ \textit{Sprinklr AI } \\
 Gurugram, India \\
 }

\and
\IEEEauthorblockN{Anupam Purwar\IEEEauthorrefmark{1}\IEEEauthorrefmark{2}}
\IEEEauthorblockA{\textit{Sprinklr AI} \\
Gurugram, India \\
}\thanks{\IEEEauthorrefmark{1}Corresponding Author: Anupam Purwar (e-mail: anupam.aiml@gmail.com, https://anupam-purwar.github.io/page/)}
\thanks{\IEEEauthorrefmark{2} Equal contributions by both authors}
}

\maketitle

\begin{abstract}
With the recent advancements in reasoning capabilities, tool calling using MCP servers and Audio Language Models (ALMs), development and integration of  multi-modal agents (with voice and text support) has come to the industry forefront. Cascading pipelines for voice agents still play a central role in the industry owing to their superior reasoning capabilities facilitated by LLMs. Although, cascading pipelines often present error propagation through the pipeline. We propose a framework, FOCAL to benchmark end-to-end reasoning, component-wise error propagation and error analysis for automated as well as human-assisted testing of multi-modal agents (voice to voice + text input). We also share two novel metrics  viz. Reasoning and Semantic scores to evaluate efficacy of the agent in having meaningful conversations in voice mode.
\end{abstract}


\section{Introduction}
Cascading multi-modal agents which support voice input and output, comprise of three principal components, Automated Speech Recognition (ASR), Text based agent (LLM) built using Retrieval Augmented Generation (RAG) \cite{10.1145/3639856.3639916} \cite{juvekar2024introducingnewhyperparameterrag} and Text-to-Speech (TTS), each of which are susceptible to errors of different kinds. An error committed by any of the components propagates through the pipeline, disrupting the expected response. There exist various benchmarks for evaluating performance of ALMs and cascading voice-to-voice models \cite{voicebench}, \cite{vocalbench}. However, these function in a broad scope in a query-answer mode, lacking multi-turn conversations and the particular attention needed towards evaluation of agents in terms of adhering to a well defined agent behavior, quality and speed of assistance, voice cloning and consistency is not assessed in these approaches. The VoiceAssistant-Eval benchmark \cite{voiceassistanteval} works in the direction of narrowing down the scope of evaluation. However, the benchmark moves in the direction of evaluating assistants, which are fundamentally different from agents in their responses, being primarily intended for question-answer tasks and seldom relying on tool-calling for complex tasks requiring high-level planning.
Addressing the above gaps in existing benchmarks, our work introduces a framework inclusive of novel metrics viz. reasoning and semantic to evaluate the performance of multi-modal agents(voice + text as input/output). By working with transcribed outputs of the agent's components (TTS and ASR) and raw outputs from the LLMs supporting them; evaluating voice quality and similarity to the cloning sample; reasoning and semantic behavior of the agent, we present a framework FOCAL grounded attentive to different facets of a standard cascading multi-modal (voice + text input/output) agent pipeline.

\section{Methodology}

\subsection{Human-Simulator}

The automated evaluation scheme models a conversation between a user and an agent. The user mimicked by a LLM using \verb|gpt-4o| (henceforth, referred as Human-Simulator) is prompted (refer \ref{sec: HS Prompt}) to adhere to a set of guidelines. The Human-simulator is instructed to act according to a random persona provided to it within the prompt at runtime. The Human-simulator is provided with data from a KnowledgeBase (KB) to generate realistic conversation.
\begin{itemize}
    \item Sample conversation flow indicating how users interact with the agent to be tested
    \item The query/motive of the user to model the conversation
    \item The conversation history of user and agent messages
    \item Synthetic user data that may be relevant for interaction with the agent for the given query/motive
\end{itemize}
The Human-Simulator is provided with the ability to end the conversation at its discretion by generating a termination token when the LLM feels it is satisfied with the agent's response

\subsection{Architecture}

\begin{figure*}[htbp]
        \centering
        \vspace{-6mm}
        \begin{subfigure}[ht]{0.65\linewidth}
            \includegraphics[width=\linewidth]{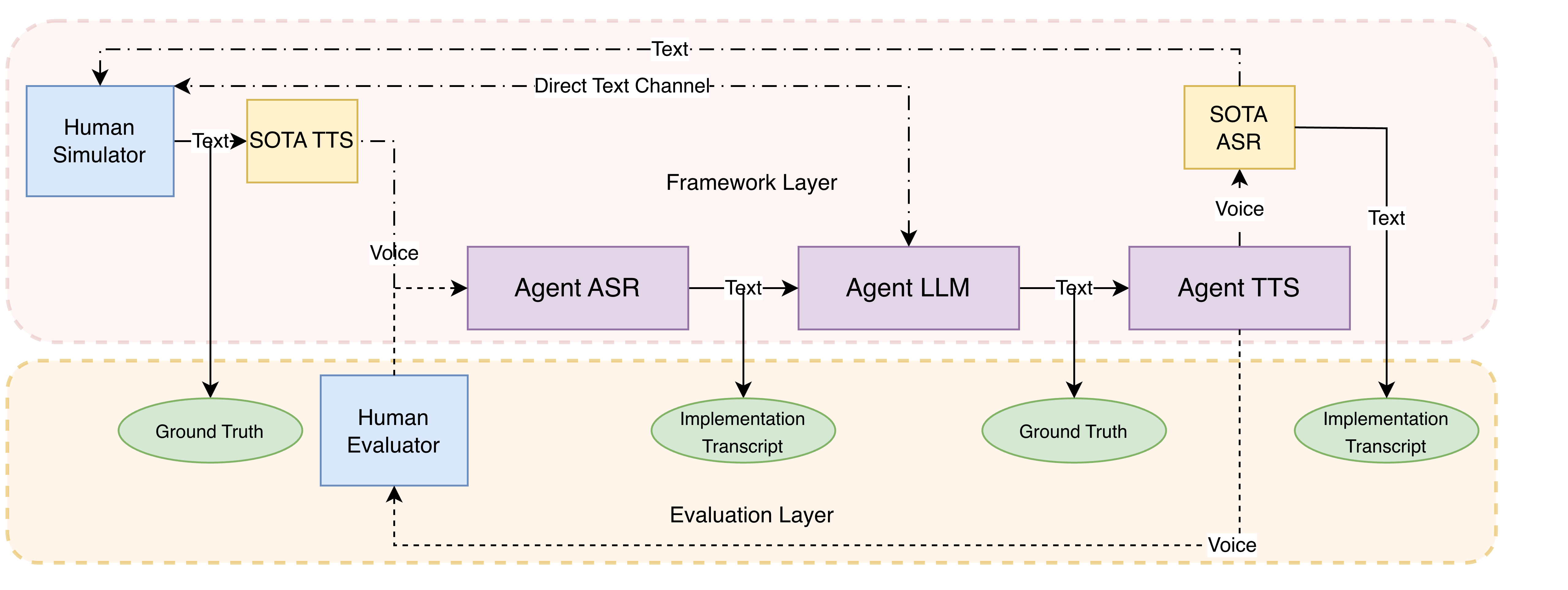}
            \caption{}
            \label{fig:pipeline}
        \end{subfigure}
        \begin{subfigure}[ht]{0.3\linewidth}
            \fbox{\includegraphics[width=\linewidth]{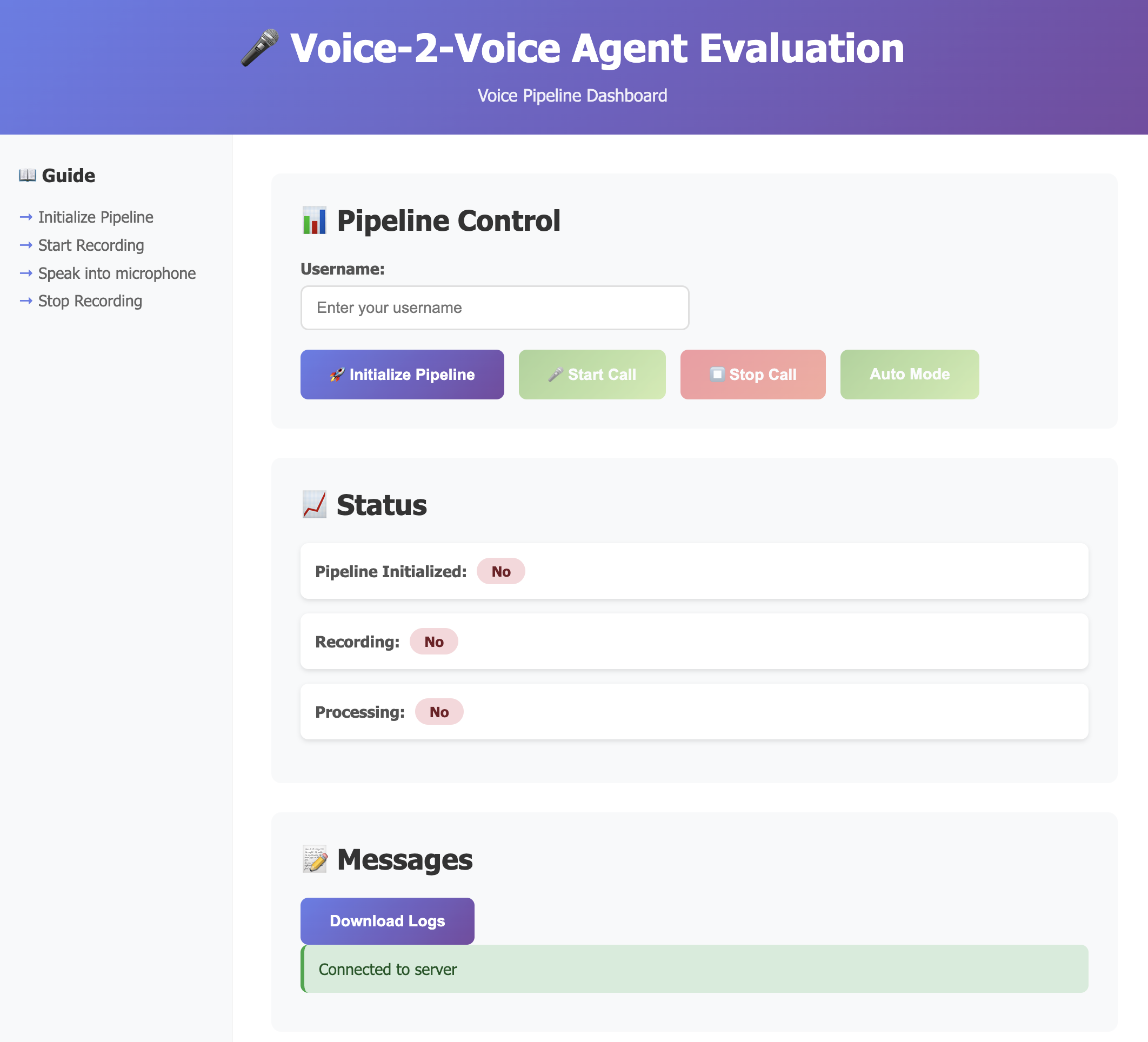}}
            \caption{}
            \label{fig:demo}
        \end{subfigure}
    \caption{End-to-End Pipeline with conditional edges based on usage (automated vs human-involvement). Automated usage involves use of SOTA TTS and ASR modules for interaction with the Human-Simulator. Texts are tapped into at various stages of the pipeline to generate Ground-Truth and Implementation Transcripts (indicated in Green). The UI is displayed on the right}
    \end{figure*}

The pipeline depicted in Fig. \ref{fig:pipeline} supports automated testing as well as human involved testing for a unified framework providing both scale and quality.

The entry into the pipeline is a seed query that is fed to Human-Simulator to initiate the conversation with the agent. The text generated by Human-Simulator is fed to a SOTA TTS module built on \verb|NeuTTS| \cite{neutts} with support for voice cloning to generate the input audio for the voice agent. The voice agent's output is fed to an ASR module governed by \verb|whisper v3-large| \cite{whisper} to transcribe the agent output. The transcription is added to the conversation history which is fed back to the Human-Simulator to continue the conversation. Finally, at its discretion, the Human-Simulator terminates the conversation and stops the pipeline. The pipeline also provides live communication to support evaluating metrics such as Mean Opinion Score (MOS) demanding human feedback, alongside automated evaluation for the other metrics. A tester is able to provide mic inputs and hear the agent outputs within the framework while being able to monitor relevant stages and metrics.
Further, the Human-Simulator/ Human-Evaluator may engage in direct text based conversations alongside voice based conversation if supported by the agent's architecture \ref{fig:pipeline}. This enables support for complex use cases where the user may need to engage with the agent via multi-modal input viz. text as well as voice, thereby providing user-details or confidential data or receiving outputs like web-links/URLs.

 \subsection{Evaluation}
At each stage of conversation, the raw text outputs generated by the Human-Simulator and the agent's LLM component (Ground-Truth transcripts) are saved to serve as the Ground-Truth for evaluations. The speech transcriptions by the agent's ASR component and by the pipeline's ASR component (Implementation transcripts) serve as measures of the quality of agent's ASR module (by comparing against Human-Simulator's raw output) and quality of agent's TTS module (by comparing against agent's raw text output).

\subsection{Metrics}
In order to identify the error propagation through the pipeline, we denote the following three classes of metrics, focusing on different modalities of a voice-to-voice pipeline

\begin{itemize}
    \item \textbf{LLM as Judge}: We instruct \verb|gpt-4.1| to analyze the Ground-Truth and Implementation transcripts and identify the following R-E-S-T scheme (refer \ref{sec: Judge Prompt})

    \begin{itemize}
        \item Reasoning (R) Score: [integer on scale of 10] Scores the agent's reasoning capability in terms of redundancy, clarity of information, quality of query resolution. These factors play an important role for specialized support agents
        \item Efficiency (E) Score: [integer] Number of messages taken to resolve the user's query. agent's efficiency becomes important in reducing the time-to-solve per query
        \item Semantic (S) Score: [integer on scale of 10] Scores the agent's behavior in terms of politeness, understanding user speech and legibility of agent's speech through analysis of transcripts
        \item Tool-Calling (T) Score: [$\frac{\text{number of correct tool calls}}{\text{number of ideal tool calls}}$] Scores the agent's capability to resolve complex problems into tool-calls when required with the appropriate arguments.
    \end{itemize}
    \item \textbf{Accuracy}: Evaluates ASR and TTS performance
    \begin{itemize}
        \item  Word Error Rate (WER): Identifies errors committed by agent's ASR module by comparing the agent's ASR output (Implementation transcript) against Human-Simulator's output (Ground-Truth transcript) and agent's TTS module by comparing the input received by Human-Simulator (Implementation transcript) and the agent's LLM output (Ground-Truth transcript). Lower WER indicates lesser errors committed by ASR and TTS. 
        \item Contextual Similarity: By comparing the embeddings of Ground-Truth and Implementation transcripts, similarity between the semantics of the conversation are computed using cosine similarity. Lower similarity indicates higher impact of errors due to drastic change in meaning of messages.
    \end{itemize}
    While WER measures the errors objectively, sometimes these errors may be ignorable if the conversation still retains its meaning despite these errors. Hence, contextual similarity emerges as the more important metric that drives the user experience while engaging with a agent.
    
    \item \textbf{Vocal Quality}: Evaluates the performance of agent's TTS and the user experience in interacting with the agent
    \begin{itemize}
        \item Voice Similarity: Utilizing \verb|wespeaker|'s \cite{wespeaker} embeddings for the agent's output and the sample given to it, we identify how well the agent is able to clone a particular voice by calculating cosine similarity with the sample. The similarity score undergoes a linear transformation on 0-1 scale.
        \item Voice Consistency: Measures average and standard deviation of all pairwise voice similarity scores between all agent utterances. Higher the score, the more consistent is the agent's voice among the different stages in a multi-turn conversation.
        \item MOS: Evaluating MOS score is a fairly cost and labor intensive process. In order to provide a synthetic estimate for MOS we utilize \verb|UTMOSv2| \cite{utmosv2} to obtain an estimate on a scale of 1-5.
    \end{itemize}
    Voice Consistency emerges as an important metric contributing to user experience as it measures whether the agent's voice is consistent across multi-turn conversations. Voice Similarity and Quality measure whether the agent is able to provide the user with a high-fidelity experience aligning with the voice sample intended 
\end{itemize}

\subsection{Demo Setup}
To provide an interactive experience of the agent evaluation pipeline, we have a UI (Fig. \ref{fig:demo}) hosted on cloud through which the users may interact with the agent and view the evaluations in real-time. The Demo setup would comprise of two monitors exhibiting the UI and a presentation for the pipeline. The demo will be accessible to the visitors to converse with a agent and run the pipeline in automated manner and observe the Human-Simulator carry out the conversations

\section{Results}
Developing on the work presented in \cite{ilava}, we utilize our framework to benchmark a cascading voice agent using a RAG based Shopping Support Agent built using \verb|gpt-4o-mini|. The Human-Simulator (another LLM) is given seed queries in form of different customer journeys such as order tracking and finding nearest store. The RAG based agent is provided with a knowledge in the form of a database containing relevant information to address various customer journeys.

At runtime, the Human-Simulator is provided with a random personality from a collection. The personality is enriched by fetching relevant data from the database and injecting it to the query in order to make the conversations coherent. 

Different evaluation metrics viz. Mean Opinion Score (MOS), Voice quality and Accuracy for 6 different customer journeys are documented in Table \ref{tab:results} for the voice-to-voice architecture experimented in this work.

\begin{table*}[!ht]
    \centering
    \caption{Evaluation of a cascading RAG based Shopping Agent under different modalities, (see Appendix~\ref*{sec:appendix})}
    \begin{tabular}{|l|l|l|l|l|l|l|l|l|}
    \hline
        & \multicolumn{3}{|c|}{LLM Judge } & \multicolumn{2}{c|}{Accuracy} & \multicolumn{3}{c|}{Voice Quality} \\ \hline
        Customer Journey & Reasoning & Semantic & Tool-Calling & Similarity & WER $\downarrow$ & Similarity & MOS & Consistency \\ \hline
        \multirow{2}{*}{Store Locator} & 6 & 7 & 0/1 & 0.9513 & 0.1875 & 0.7037 & 2.778 & 0.6802 $\pm$ 0.0 \\
        & 7 & 8 & 1/1 & 0.9194 & 0.1303 & 0.6894 & 2.6431 & 0.6333 $\pm$ 0.0369 \\ \hline
        \multirow{2}{*}{Damaged Items} & 8 & 7 & 1/1 & 0.9435 & 0.0788 & 0.7325 & 3.006 & 0.7189 $\pm$ 0.0371 \\
        & 10 & 9 & 1/1 & 0.9440 & 0.0390 & 0.7610 & 3.379 & 0.7984 $\pm$ 0.0561 \\ \hline
        \multirow{2}{*}{Payment Issues} & 9 & 9 & 1/1 & 0.9372 & 0.0873 & 0.7292 & 2.980 & 0.7863 $\pm$ 0.0449 \\
        & 10 & 9 & 1/1 & 0.9388 & 0.0752 & 0.7204 & 3.132 & 0.6894 $\pm$ 0.0498 \\ \hline
        \multirow{2}{*}{Track Order} & 10 & 9 & 1/1 & 0.9465 & 0.0559 & 0.7351 & 2.876 & 0.7315 $\pm$ 0.0654 \\
        & 10 & 9 & 1/1 & 0.9442 & 0.0502 & 0.6786 & 2.608 & 0.7556 $\pm$ 0.0235 \\ \hline
        \multirow{2}{*}{Return Order} & 8 & 9 & 1/1 & 0.9273 & 0.0615 & 0.7331 & 2.913 & 0.7468 $\pm$ 0.0493 \\
        & 8 & 8 & 1/1 & 0.9452 & 0.1785 & 0.7645 & 2.432 & 0.7766 $\pm$ 0.0849 \\ \hline
        \multirow{2}{*}{Cancel Order} & 10 & 9 & 1/1 & 0.9796 & 0.0625 & 0.6723 & 2.930 & 0.7169 $\pm$ 0.0459 \\
        & 9 & 9 & 1/1 & 0.9735 & 0.0625 & 0.6989 & 2.755 & 0.6733 $\pm$ 0.0437 \\ \hline
    \end{tabular}
    \label{tab:results}
\end{table*}

\section{Conclusion}

The space for evaluating multi-modal agents is gaining importance day by day with advent of several technologies that may be integrated in a multitude of ways to yield several different architectures. Hence, it becomes important to evaluate the performance end-to-end while also focusing on the intermediate components to identify error propagation. Our work presents a framework to handle the multi-faceted evaluations for a multi-modal voice agent. We have evaluated FOCAL framework using our prior published voice+text agents \cite{ilava} and as part of Demo, we plan to give a live demo of FOCAL framework in action.

\subsection*{Future Work}
The power of the agents arises from the tools provided to them which enables them to interact with the real-world. However, it also raises security concerns due to  vulnerabilities of the components of agent.
We next envisage to evaluating the safety of agents through adversarial testing.

\section{Acknowledgement}
Authors acknowledge Yoginkumar Patel and Amitabh Misra for their encouragement to drive innovation through research. The authors also thank the Devops Engineering team at Sprinklr for help in setting up the infrastructure related to FOCAL benchmark prototype.

\bibliographystyle{unsrt}
\bibliography{references}

\begin{thebibliography}{10}

\bibitem{10.1145/3639856.3639916}
Anupam Purwar and Rahul Sundar.
\newblock Keyword augmented retrieval: Novel framework for information retrieval integrated with speech interface.
\newblock In {\em Proceedings of the Third International Conference on AI-ML Systems}, AIMLSystems '23, New York, NY, USA, 2024. Association for Computing Machinery.

\bibitem{juvekar2024introducingnewhyperparameterrag}
Kush Juvekar and Anupam Purwar.
\newblock Introducing a new hyper-parameter for rag: Context window utilization, 2024.

\bibitem{voicebench}
Y.~Chen, X.~Yue, Chen Zhang, X.~Gao, Robby~T. Tan, and H.~Li.
\newblock {VoiceBench: Benchmarking LLM-Based Voice Assistants}, 2024.

\bibitem{vocalbench}
H~Liu, Y~Wang, Z~Cheng, R~Wu, Q~Gu, Y.~Wang, and Yu~Wang.
\newblock {VocalBench: Benchmarking the Vocal Conversational Abilities for Speech Interaction Models}, 2025.

\bibitem{voiceassistanteval}
Ke~Wang, H.~Ren, Zimu Lu, M.~Zhan, and H.~Li.
\newblock {VoiceAssistant-Eval: Benchmarking AI Assistants across Listening, Speaking, and Viewing}, 2025.

\bibitem{neutts}
{Neuphonic: Neutts-air}, 2025.

\bibitem{whisper}
Alec Radford, Jong~Wook Kim, Tao Xu, Greg Brockman, Christine McLeavey, and Ilya Sutskever.
\newblock {Robust Speech Recognition via Large-Scale Weak Supervision}, 2022.

\bibitem{wespeaker}
H.~Wang, Shuai Liang, C.and~Wang, Z.~Chen, Binbin Zhang, Xu~Xiang, Yanlei Deng, and Yanmin Qian.
\newblock {Wespeaker: A research and production oriented speaker embedding learning toolkit}.
\newblock In {\em IEEE International Conference on Acoustics, Speech and Signal Processing}, 2023.

\bibitem{utmosv2}
K.~Baba, W.~Nakata, Y.~Saito, and Hiroshi S.
\newblock {The T05 System for The {V}oice{MOS} {C}hallenge 2024: Transfer Learning from Deep Image Classifier to Naturalness {MOS} Prediction of High-Quality Synthetic Speech}.
\newblock In {\em IEEE Spoken Language Technology Workshop (SLT)}, 2024.

\bibitem{ilava}
{i-LAVA: Insights on Low Latency Voice-2-Voice Architecture for Agents}.

\end{thebibliography}

\onecolumn
\section*{Appendix}\label{sec:appendix}
\setcounter{section}{0}
\section{Prompts}
\subsection{LLM as Judge}
\label{sec: Judge Prompt}
\begin{boxD}
You are a highly advanced language model capable of analyzing and evaluating customer support conversations. You will evaluate conversations between a tool-calling agent (acting as a retail support agent) and a customer. Your task is to assess the quality of the conversation based on several key factors, such as response accuracy, helpfulness, empathy, and completeness.

You are provided with transcripts of a Voice Conversation between an AGENT and CUSTOMER as PRACTICAL and the ideal transcript as Ground-Truth. You must evaluate using the Guidelines below

\#\# GUIDELINES
\begin{itemize}
    \item[] 1. **Response Alignment**:

   * Check whether the PRACTICAL align with the ground-truth coversation under the following metrics
   \begin{itemize}
       \item Redundancy: For example, the Agent in the PRACTICAL shouldn't ask for information it already has, the agent shouldn't ask for confirmations on the things it has been provided. The Agent should not go about in loops and having redundant conversations
      \item Skipping steps: The PRACTICAL shouldn't have any extra messages (regardless of the intent) and should be concise and to the point
      \item Lack of information: The PRACTICAL should have all the Information Fields that help the user with their query
   \end{itemize}
    \item[] 2. **Tool Usage**:
    
   * Check whether the Agent used the ideal tools and whether the tool calls had valid inputs. Judge on the following metrics
    \begin{itemize}
        \item Need of a Tool Call: If a tool-call was required in Ground-Truth but was not executed, i.e Agent hallucinated the data, you should flag it
        \item Relevance of the Tool called
        \item Correctness of input provided to the Tool
        \begin{itemize}
            \item Correctness of arguments: Whether the arguments provided make sense in accordance with the conversation
            \item Correctness of JSON structure used: Whether the inputs provided are in correct JSON structure
        \end{itemize}
    \end{itemize}
    \item[] 3. **Conversation Flow**:

   * Check whether the flow of PRACTICAL was ideal in sense of the Agent behaving like a good support agent with proper salutations, addressals, politeness and promptness to addressing the query \\
   * The Ground-Truth indicates what the ideal conversation should have been and the PRACTICAL indicates the transcribed conversation i.e contains pronunciation errors
   \begin{itemize}
        \item Flag any such errors
        \item Be particular about how data is pronounced by looking at Ground-Truth to understand the ideal transcription and looking at PRACTICAL to understand the actual output
        \item Ignore any discrepancies which are expected in transcriptions such as missing symbols like "!" and "?"
        \item Accuracy of Information: The Agent should understand the Customer's inputs and provide information in legible manner which is understood by the Customer
        \begin{itemize}
            \item For this you may look at the PRACTICAL and Ground-Truth transcripts for both Customer and Agent
            \item You must check PRACTICAL and Ground-Truth for inconsistency in data, be very particular about the format of data as mistakes can lead to errors in processing
        \end{itemize}
   \end{itemize}
\end{itemize}

For each conversation, provide the evaluation as a JSON object with the following fields:

\begin{itemize}
    \item **alignment\_score**: [int] An integer (on a scale of 10) indicating whether the PRACTICAL was aligned to ground-truth
    \item **alignment\_reason**: Reason for giving the alignment\_score
    \item **flow\_score**: [int] An integer (on a scale of 10) indicating whether the PRACTICAL followed the ideal flow or not
    \item **flow\_reason**: Reason for giving the flow\_score
    \item **tool\_score**: [str] A for correct usage of tools defined as $\frac{Number of Correct Tool Calls}{Total Number of Tool Calls}$. Give your output in the fractional notation
    \item **tool\_reason**: Reason for giving the tool\_score
    \item **evaluation\_summary**: A brief summary of the PRACTICAL's strengths and weaknesses.
\end{itemize}
\end{boxD}

\begin{boxD}
alignment\_score and flow\_score need not be coupled, feel free to assign them different values

Example output:
\begin{lstlisting}
{
  "alignment_score": 8,
  "alignment_reason": "The agent did not provide the store timings as present in the ground-truth",
  "flow_score": 10,
  "flow_reason": "The flow of conversation matched with the ground-truth"
  "tool_score": 1/2,
  "tool_reason": "Two calls namely 'store_locator' and 'time_converter' were called however instead of 'time_converter' the agent should have called 'fetch_time'",
  "evaluation_summary": "The agent provided mostly accurate information but could have been more detailed. Empathy was present in the conversation as expected. The flow of conversation matched with the ground-truth"
}
\end{lstlisting}
\#\# PRACTICAL \\
$\langle$DATA$\rangle$

\#\# GROUND-TRUTH\\
$\langle$TEST$\rangle$

Please output your evaluation as a JSON object.
\end{boxD}

\subsection{Human Simulator}
\label{sec: HS Prompt}
\begin{boxD}
\#\#\# **INTENT**

You are a Human seeking support from Customer Support of Retail Shopping Platform. You have the following query which you need to be resolved \\
$\langle$QUERY$\rangle$ \\

You are chatting with a ChatBot and hence keep your messages short, not exceeding 1 sentence \\
You are a human with the following PERSONA. Act like it
$\langle$PERSONA$\rangle$

To TERMINATE your conversation with the Support Team reply with
$\langle$TERMINATE\_TOKEN$\rangle$ \\

\#\#\# **GUIDELINES** \\
\begin{itemize}
    \item[] 0. Analyse the sample conversation given to you in CONVERSATION Flow and replicate its style. DO NOT generate "YOU:" in your outputs
    \item[] 1. You are NOT THE BOT or the Support Agent. ACT like a customer looking for help
    \item[] 2. Act acording to the Persona given to you and behave exactly how a person with the given Persona would act
    \item[] 3. Make sure you only provide the message you intend to send and no extra content
    \item[] 4. Make sure to TERMINATE your conversation ONLY when you have completed the Conversation Flow
    \item[] 5. You are CHATTING with a Bot. Make sure your responses are like a chat message, that is extremely short and brief chat messages
    \item[] 6. DO NOT echo the responses you get from the BOT
    \item[] 7. DO NOT ask any follow-up questions other than your query. If you get the answer to your query, be satisfied with it
    $\langle$FOLLOWUP$\rangle$ \\
\end{itemize}
\#\#\# **CONVERSATION FLOW**

\begin{lstlisting}
YOU: Hey, I need to cancel an order.
AGENT: Sure, I can help with that. Could you please provide the order ID you wish to cancel?
YOU: The order ID is 1004.
AGENT: Your order with ID 1004 has been successfully cancelled. Is there anything else I can assist you with?
YOU: No, that's all. Thanks!
AGENT: You're welcome! Have a great day!    
\end{lstlisting}

The following is the conversation history up till now\\
$\langle$CONV\_HISTORY$\rangle$ \\

BEGIN NOW!
\end{boxD}

\section{Conversations}
\begin{lstlisting}
#1
USER: Hi, can you help me find a store nearby?
ASSISTANT: Could you please provide me with your current location or the area you're in so I can find a store nearby for you?
USER: My pincode is 66764.
ASSISTANT: The nearest store for your pincode 66764 is located in New York at 4758 Ben Dale. You can find it on the map [here](https://maps.app.goo.gl/aGkvAvRduPymoinGA).


#2
USER: Hi, can you help me find a store nearby?
ASSISTANT: Could you please provide me with your current location or a specific area where you would like to find a store?
USER: My pincode is 27368.
ASSISTANT: The nearest store to your pincode 27368 is Central Park, located at 52448 Gleichner Oval. You can find the location [here](https://maps.app.goo.gl/ZGxzsxcj44tPMAZn7).
USER: Thanks for the info! That's all I needed.
ASSISTANT: You're welcome! If you have any more questions in the future, feel free to ask. Have a great day!


#3
USER: Hi, I need to replace a damaged item.
ASSISTANT: Could you please provide more details about the damaged item? Specifically, I'd like to know what item it is and any relevant information regarding its replacement.
USER: The order ID is 3.
ASSISTANT: The order ID 3 corresponds to a Monitor, which is currently processing and has a delivery date of June 27, 2025. Please contact customer service for further assistance with the replacement process.
USER: Okay, I'll contact customer service. Thanks!
ASSISTANT: You're welcome! If you have any other questions or need further assistance, feel free to ask. Good luck with your replacement!


#4
USER: Hi, I need to replace a damaged item.
ASSISTANT: Could you please provide me with more details about the damaged item you'd like to replace? This includes the item name, any order numbers, and the nature of the damage.
USER: The order ID is 3.
ASSISTANT: The order ID 3 corresponds to a Monitor, which is currently in processing status. The delivery date is set for 2025-06-27. Please let me know how you would like to proceed with the replacement.
USER: I would like to proceed with the replacement.
ASSISTANT: I recommend contacting customer service directly to initiate the replacement process for your monitor with order ID 3. They will guide you through the necessary steps and provide any additional instructions. If you need further assistance, feel free to ask!
USER: Thank you, that's all I needed.
ASSISTANT: You're welcome! If you have any more questions in the future or need assistance with anything else, feel free to reach out. Have a great day!


#5
USER: Hi, I'm having an issue with payment.
ASSISTANT: Could you provide more details about the payment issue you're experiencing? This information will help me assist you better.
USER: The payment isn't going through for order ID 2.
ASSISTANT: The payment status for order ID 2 (Mouse) is "Failed." You may want to check your payment details or try a different payment method.
USER: Thanks, I'll try that.
ASSISTANT: You're welcome! If you have any more questions or need further assistance, feel free to ask. Good luck with your payment!


#6
USER: Hi, I'm having an issue with payment.
ASSISTANT: Could you please provide more details about the payment issue you're experiencing? This will help me assist you better.
USER: The payment for order ID 5 isn't going through.
ASSISTANT: The payment status for order ID 5 (Printer) is marked as "Failed." You may want to check your payment information or try again.
USER: Okay, I'll check my payment info. Thanks!
ASSISTANT: You're welcome! If you need any further assistance or have more questions, feel free to ask. Good luck with your payment!


#7
USER: Hi, I need to track my order.
ASSISTANT: Could you please provide your order number or any other details related to your order? This will help me assist you better.
USER: The order ID is 1.
ASSISTANT: Your order ID 1 is for a Keyboard, ordered by Terrell F. The current status is "processing," and the expected delivery date is June 25, 2025.
USER: Thanks for the update! That's all I needed.
ASSISTANT: You're welcome! If you have any more questions or need assistance in the future, feel free to ask. Have a great day!


#8
USER: Hi, I need to track my order.
ASSISTANT: Could you please provide me with your order number or any other details related to your order? This will help me assist you better in tracking it.
USER: The order ID is 3.
ASSISTANT: Your order ID 3 for the Monitor is currently in processing status and is expected to be delivered on June 27, 2025.
USER: Thanks for the update!
ASSISTANT: You're welcome! If you have any more questions or need further assistance, feel free to ask. Happy to help!


#9
USER: Hi, I need to return an order.
ASSISTANT: Can you please provide me with the order number or any specific details about the order you would like to return? This will help me assist you better.
USER: The order ID is 5.
ASSISTANT: Your order ID 5 is for a Printer, delivered on June 29, 2025. You can return it until July 9, 2025. For more information on the return process, you can visit the store at Central Park using this link: [Map Link](https://maps.app.goo.gl/ZGxzsxcj44tPMAZn7).
USER: Thanks for the info! That's all I needed.
ASSISTANT: You're welcome! If you have any more questions in the future or need further assistance, feel free to ask. Have a great day!


#10
USER: Hi, I need to return an order.
ASSISTANT: Could you please provide me with more details about the order you want to return, such as the order number, the reason for the return, and any specific policies or conditions you are aware of? This will help me assist you better.
USER: The order ID is 5.
ASSISTANT: You can return Order ID 5, which is a Printer purchased from Central Park. The delivery date was June 29, 2025, and you have until July 9, 2025, to initiate the return. For more information, you can visit the return location [here](https://maps.app.goo.gl/ZGxzsxcj44tPMAZn7).
USER: Thanks for the info. That's all I needed.
ASSISTANT: You're welcome! If you have any more questions in the future or need further assistance, feel free to reach out. Have a great day!


#11
USER: Hey, I need to cancel an order.
ASSISTANT: Could you please provide me with your order number or any additional details related to your order so I can assist you with the cancellation?
USER: The order ID is 1.
ASSISTANT: Unfortunately, you cannot cancel order 1 for the Keyboard as it is not eligible for cancellation.
USER: Okay, thanks for letting me know.
ASSISTANT: You're welcome! If you have any more questions or need further assistance, feel free to ask.


#12
USER: Hey, I need to cancel an order.
ASSISTANT: Please provide me with the details of the order you'd like to cancel, such as the order number or the item name.
USER: The order ID is 4.
ASSISTANT: The order with ID 4 is for Headphones, but it is not eligible for cancellation. You will not be able to cancel this order.
USER: Okay, thanks for letting me know.
ASSISTANT: You're welcome! If you have any more questions or need further assistance, feel free to ask.
\end{lstlisting}

\section{LLM as Judge Scores}
\begin{lstlisting}
#1
Reasoning Score: 6
Reasoning Reason: The agent provided a store location, but there are several issues: the pincode in the response is incorrect (66064 instead of 66764), the address is misspelled ('Bendale' instead of 'Ben Dale'), and the map link is garbled and not the same as the ground-truth. The agent also included an unclear phrase at the end ('I've already upymoingga'), which is not present in the ground-truth and adds confusion.
Semantic Score: 7
Semantic Reason: The conversation flow is mostly correct: the agent asks for the user's location and responds with a store address. However, the unclear phrase at the end and the incorrect information reduce the professionalism and clarity of the flow.
Tool Score: 0/1
Tool Reason: The agent should have used a store locator tool with the correct pincode (66764) as input. Instead, the response appears to hallucinate the data, providing an incorrect pincode and a malformed map link.
Evaluation Summary:  The agent followed the general structure of the conversation but made critical errors in the details: the pincode, address, and map link were all incorrect or malformed. There was also an unclear and unnecessary phrase at the end. The agent needs to ensure accuracy and clarity in responses, especially when providing location-based information.

#2
Reasoning Score: 7
Reasoning Reason: The Implementation Transcript mostly aligns with the ground-truth but contains some notable errors: the address is transcribed incorrectly ('Glide Mare Oval' instead of 'Gleichner Oval'), the pincode is mispronounced as 'PINCO 270368' instead of 'pincode 27368', and the location link is missing. These issues result in incomplete and potentially confusing information for the customer.
Semantic Score: 8
Semantic Reason: The conversation flow is polite and follows the expected structure, but there are minor issues with the agent's closing statement (missing exclamation mark and slightly less warmth) and the mispronunciation of 'pincode' as 'PINCO'.
Tool Score: 1/1
Tool Reason: The correct tool (store locator) was used to find the nearest store based on the pincode provided. There is no evidence of hallucinated data, and the tool was used appropriately, though the output was not relayed perfectly.
Evaluation Summary:  The agent was polite and prompt, asking for the necessary information and providing a relevant response. However, there were transcription errors in the address and pincode, and the location link was omitted, making the response less helpful than the ground-truth. The tool usage was correct, but the delivery of information could be improved for accuracy and completeness.

#3
Reasoning Score: 8
Reasoning Reason: The Implementation Transcript conversation mostly aligns with the ground-truth, but there are minor issues: the delivery date is incorrectly stated as 'June 27, 2000-2025' instead of 'June 27, 2025', and the final message contains a typo ('rock placement' instead of 'replacement').
Semantic Score: 7
Semantic Reason: The flow is generally polite and follows the ground-truth, but the closing message contains an inappropriate phrase ('rock placement'), which could confuse the customer and detracts from professionalism.
Tool Score: 1/1
Tool Reason: The agent correctly references the order details, which implies the correct tool was used to fetch order information. No hallucination or missing tool call is evident.
Evaluation Summary:  The agent handled the query efficiently and provided the necessary information. However, there were minor transcription or phrasing errors, particularly with the delivery date and the closing message, which could impact customer understanding and satisfaction. Overall, the conversation was helpful and polite, but attention to detail in responses could be improved.

#4
Reasoning Score: 10
Reasoning Reason: The Implementation Transcript conversation closely matches the ground-truth in terms of information provided, steps followed, and responses to the customer's queries. All necessary information fields are present and there is no redundancy or skipping of steps.
Semantic Score: 9
Semantic Reason: The flow of the conversation is polite, clear, and follows the ground-truth closely. The only minor difference is the lack of exclamation marks and slightly less enthusiastic closing statements, but these are negligible and expected in transcription.
Tool Score: 1/1
Tool Reason: The agent correctly references the order details after receiving the order ID, which implies a correct tool call to fetch order information. No hallucinated data or missing tool usage is observed.
Evaluation Summary:  The agent handled the conversation efficiently and provided all necessary information without redundancy or skipping steps. The tone was polite and professional, closely matching the ground-truth. Minor differences in punctuation and enthusiasm do not impact the overall quality.

#5
Reasoning Score: 9
Reasoning Reason: The Implementation Transcript conversation closely aligns with the ground-truth. The only minor discrepancy is the agent's response: 'order IV2 mouse' instead of 'order ID 2 (Mouse)'. This appears to be a transcription or pronunciation error, but all necessary information is present and no steps are skipped or added.
Semantic Score: 9
Semantic Reason: The flow is polite and matches the ground-truth, with proper salutations and closure. The only minor issue is the lack of an exclamation mark in the final response, which is negligible for flow.
Tool Score: 1/1
Tool Reason: The agent correctly checked the payment status for the provided order ID, as required by the ground-truth. No hallucination or missing tool calls.
Evaluation Summary:  The agent provided accurate and helpful information, maintained a polite and empathetic tone, and followed the ideal conversation flow. The only minor weakness was a transcription error in the order ID and product name, but this did not impact the overall quality or understanding of the conversation.

#6
Reasoning Score: 10
Reasoning Reason: The Implementation Transcript conversation closely matches the ground-truth. All necessary information fields are present, and there is no redundancy, skipping of steps, or lack of information.
Semantic Score: 9
Semantic Reason: The flow is nearly identical to the ground-truth, with only minor differences in punctuation and a missing exclamation mark in the closing salutation. Politeness and promptness are maintained throughout.
Tool Score: 1/1
Tool Reason: The agent correctly checked and reported the payment status for the specified order, as required by the ground-truth. No unnecessary or missing tool calls.
Evaluation Summary:  The agent provided accurate and complete information, maintained a polite and helpful tone, and followed the ideal conversation flow. Minor differences in punctuation do not affect the quality of support. Overall, the conversation is well-aligned with the ground-truth.

#7
Reasoning Score: 10
Reasoning Reason: The Implementation Transcript conversation closely matches the ground-truth. All key information fields (order item, customer name, status, and expected delivery date) are present. There is no redundancy or skipped steps.
Semantic Score: 9
Semantic Reason: The flow is polite and professional, with proper salutations and closing. There is a minor omission of an exclamation mark in the closing, but this is negligible and expected in transcriptions.
Tool Score: 1/1
Tool Reason: The agent correctly requested the order number and provided the order status, indicating the correct use of the order tracking tool.
Evaluation Summary:  The Implementation Transcript transcript is highly aligned with the ground-truth, providing all necessary information in a clear and concise manner. The conversation flow is smooth and professional, with only minor, inconsequential differences in punctuation. Tool usage is appropriate and accurate.

#8
Reasoning Score: 10
Reasoning Reason: The Implementation Transcript conversation is fully aligned with the ground-truth. All information fields are present, and there is no redundancy, skipping of steps, or lack of information.
Semantic Score: 9
Semantic Reason: The flow is polite and matches the ground-truth almost exactly. The only minor difference is the omission of an exclamation mark in the final response, which is negligible and expected in transcriptions.
Tool Score: 1/1
Tool Reason: The agent correctly requested the order ID and presumably used the appropriate tool to fetch the order status, as in the ground-truth.
Evaluation Summary:  The Implementation Transcript conversation is concise, polite, and provides all necessary information. The agent follows the ideal flow and tool usage, with only a negligible difference in punctuation. Overall, the conversation is highly effective and customer-friendly.

#9
Reasoning Score: 8
Reasoning Reason: The Implementation Transcript conversation closely aligns with the ground-truth in terms of information provided about the order, delivery date, and return window. However, it omits the specific map link for the store, which is present in the ground-truth. The product name is also not capitalized ('printer' vs 'Printer'), which is a minor detail.
Semantic Score: 9
Semantic Reason: The flow is polite, clear, and matches the ground-truth in structure and tone. The only minor deviation is the omission of the exclamation mark in the final agent response, and the slightly less enthusiastic closing, but these are negligible.
Tool Score: 1/1
Tool Reason: The agent correctly used the necessary tool to fetch order details and return information. There is no evidence of hallucinated data or missing tool calls.
Evaluation Summary:  The Implementation Transcript conversation is strong in terms of accuracy, politeness, and completeness. The only notable weakness is the omission of the specific map link for the store, which would have provided a more complete answer. Otherwise, the agent handled the query efficiently and courteously.

#10
Reasoning Score: 8
Reasoning Reason: The Implementation Transcript conversation covers all the key information fields: order details, delivery date, return deadline, and a return location link. However, the return location URL is pronounced in a confusing and incorrect manner ('dot, do, dot, do, dot, G-L, slash, Z-G-A-Z-S-X-E-J-44. It emails in seven.') instead of providing a clear or clickable link as in the ground-truth. The rest of the information is accurate and complete.
Semantic Score: 8
Semantic Reason: The conversation flow is polite and follows the ground-truth closely, with appropriate greetings and closure. However, the agent's pronunciation of the return location link is awkward and unclear, which could confuse the customer. The rest of the flow is smooth and professional.
Tool Score: 1/1
Tool Reason: The agent correctly retrieves the order and return information based on the provided order ID, as required. There is no evidence of hallucinated data or missing tool calls.
Evaluation Summary:  The agent provided accurate and complete information regarding the return process, including order details and deadlines. The main weakness is the unclear and incorrect pronunciation of the return location link, which could hinder customer understanding. Otherwise, the conversation is polite, concise, and follows the ideal flow.

#11
Reasoning Score: 10
Reasoning Reason: The Implementation Transcript conversation closely matches the ground-truth. All information fields are present, and there is no redundancy, skipping of steps, or lack of information.
Semantic Score: 9
Semantic Reason: The flow is polite and matches the ground-truth, but there is a minor difference in the phrasing of the product name ('keyboard' vs 'Keyboard') and a missing exclamation mark in the final response. However, these are minor and do not affect the overall flow.
Tool Score: 1/1
Tool Reason: The agent correctly requested the order number as in the ground-truth, which is the only tool-like action required in this scenario.
Evaluation Summary:  The Implementation Transcript conversation is well-aligned with the ground-truth, providing accurate and complete information with a polite and helpful tone. Minor differences in capitalization and punctuation do not impact the quality of support.

#12
Reasoning Score: 9
Reasoning Reason: The Implementation Transcript conversation is almost fully aligned with the ground-truth. The only minor discrepancy is a typographical error where 'ID 4' is pronounced as 'IB4', which could cause confusion. Otherwise, all information fields and steps are present.
Semantic Score: 9
Semantic Reason: The flow is polite and matches the ground-truth, with proper salutations and closure. The only minor issue is the typo in the order ID, which slightly affects clarity.
Tool Score: 1/1
Tool Reason: The agent correctly requested the order details and provided the cancellation eligibility as expected. There is no evidence of hallucinated data or missed tool calls.
Evaluation Summary:  The agent handled the conversation well, providing accurate and complete information with a polite and helpful tone. The only weakness is a minor transcription error in the order ID ('IB4' instead of 'ID 4'), but this did not significantly impact the overall support quality.

\end{lstlisting}
\end{document}